\documentclass[twocolumn,aps,prl]{revtex4}
\usepackage{graphicx}
\usepackage{dcolumn}
\begin{document}
\hspace{-0.38cm}{\bf Comment on: "Ferroelectricity in spiral magnets"}\\
\hspace{1cm}

There is much interest in the physics of materials that show a
strong coupling between magnetic and electric degrees of freedom. In
a recent paper by Mostovoy \cite{Mostovoy}, a theory is presented
that is based on symmetry arguments and leads to quite general
claims which we feel merit some further analysis. In particular,
Mostovoy concludes that {\it spiral magnets are, in general,
ferroelectric}.\par

We argue that this conclusion is not generally valid, and that the
symmetry of the unit cell has to be taken into account by any
symmetry-based magneto-electric coupling theory. In an attempt to
avoid further confusion in the search of new multiferroic materials,
we identify in this Comment some of the necessary symmetry
properties of spiral magnets that can lead to ferroelectricity.\par

\begin{figure}[!b]
\begin{center}
  \includegraphics[height=6cm,bbllx=50,bblly=307,bburx=590,
  bbury=725,angle=0,clip=]{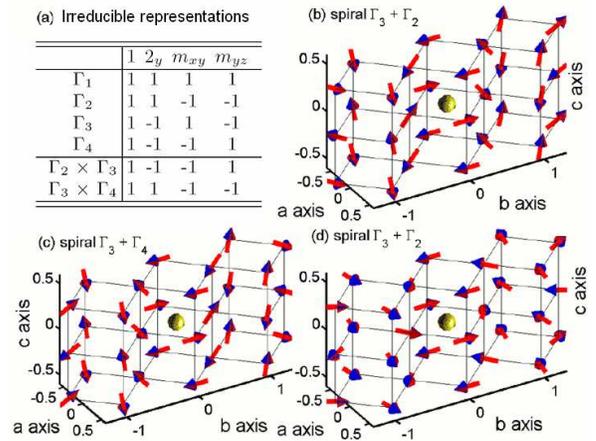}
  \caption{(a) Magnetic irreducible
  representations of the ${\rm Mn^{3+}}$ sites in ${\rm TbMnO_3}$.
  (b) Inversion-symmetry breaking spiral phase
  that was found experimentally and leads to ferroelectricity. (c)
  Hypothetical inversion-symmetry breaking spiral structure that does
  not allow ferroelectricity (the yellow sphere indicates the broken
  inversion center). (d) Hypothetical inversion-symmetry breaking spiral
  structure that allows ${\bf P} || {\bf c}$, but for which Mostovoy's
  theory predicts ${\bf P} || {\bf a}$.}
  \label{Fig1}
\end{center}
\end{figure}

We take the example of the ferroelectric phase of ${\rm TbMnO_3}$,
where the magnetic structure is incommensurate along the
crystallographic b-axis, and contains ${\rm Mn^{3+}}$ moments along
the b-axis and c-axis that belong to two different irreducible
representations $\Gamma_3$ and $\Gamma_2$, respectively
\cite{KenzelmannPRL}. The symmetry of the magnetic structure is
described by the direct product of $\Gamma_3$ and $\Gamma_2$.
Fig.~\ref{Fig1}a shows that this magnetic structure breaks inversion
symmetry {\it and} a mirror plane in the ab plane, thereby allowing
a ferroelectric polarization ${\bf P} || {\bf c}$ (but not along
other directions) \cite{KenzelmannPRL}. We reached similar
conclusions for multiferroic ${\rm Ni_3V_2O_8}$
\cite{Lawes,KenzelmannPRB}.\par

Now consider a hypothetical magnetic structure for ${\rm TbMnO_3}$
that is a magnetic spiral with ${\rm Mn^{3+}}$ moments still in the
crystallographic bc-plane, but with the c spin component belonging
to $\Gamma_4$ instead $\Gamma_2$. This structure is also a
bc-polarized spiral structure, albeit with different phase relations
between some of the nearest-neighbors compared to those in the
experimentally observed magnetic structure. The symmetry of this
structure is given by the direct product of $\Gamma_3$ and
$\Gamma_4$. Since the relevant magnetic structure is odd under both
$m_{xy}$ and $m_{yz}$ and even under $2_y$ it can not support a
polar axis: the ferroelectric moment must be zero even though
inversion symmetry is broken. In contrast, according to Eq.~5 of
Mostovoy \cite{Mostovoy}, his theory would predict a ferroelectric
moment along the c-axis.\par

%

Mostovoy's theory may also wrongly predict the direction of electric
polarization in some materials. Consider a hypothetical magnetic
structure that is a spiral modulated along the b-axis, and where the
moments along the b-axis belong to $\Gamma_3$ and the moments along
the a-axis belong to $\Gamma_2$. According to Mostovoy's theory,
this spiral structure supports ${\bf P} || {\bf a}$. However, as
shown above, a magnetic spiral structure belonging to $\Gamma_3$ and
$\Gamma_2$ can only support ${\bf P} || {\bf c}$, irrespective of
the plane of spin rotation.\par

In conclusion, we do not agree with Mostovy's statement (which most
of his paper is based on) that {\it incommensurate spin-density-wave
states are largely insensitive to details of crystal structure and
can be described by a continuum field theory of the Ginzburg-Landau
type}. The examples that we present in this Comment show that the
symmetry properties of the crystal lattice and of the magnetic order
play a crucial role in phenomenological description of
magneto-electric coupling \cite{KenzelmannPRL,Lawes}, and that the
continuum symmetry approach that Mostovoy proposes leads to
misleading predictions. Further we note that an additional virtue of
dealing with the symmetry of representations
\cite{KenzelmannPRL,Lawes,KenzelmannPRB} is that one sees
immediately that perturbations within the representations that lead
to magnetic components in addition to either the nonferroelectric
collinear structure or to the ferroelectric spiral do not change the
symmetry. To reach the same conclusion within Mostovoy's formulation
requires additional analysis.\par \vspace{0.3cm}


M. Kenzelmann\\
Laboratory for Solid State Physics, ETH Z\"{u}rich, CH-8093
Z\"{u}rich, Switzerland \\Laboratory for Neutron Scattering, ETH
Z\"{u}rich \& Paul Scherrer Institute, CH-5232 Villigen,
Switzerland\par A.~B. Harris\\Department of Physics and Astronomy,
University of Pennsylvania, Philadelphia, Pennsylvania 19104,
USA\par


\end{document}